# The coordination dynamics of social neuromarkers

E. Tognoli[1*], J. A. S. Kelso[1,2]

[1] The Human Brain and Behavior Laboratory, Center for Complex Systems and Brain Sciences, Florida Atlantic University, Boca Raton, FL, USA

[2] Intelligent System Research Centre, University of Ulster, Derry, N. Ireland

**Correspondence**:

Emmanuelle Tognoli
Center for Complex Systems and Brain Sciences
Florida Atlantic University
777 Glades Road
Boca Raton, FL-33431, USA
tognoli@ccs.fau.edu

**Abstract**

Social behavior is a complex integrative function that entails many aspects of the brain's sensory, cognitive, emotional and motor capacities. Neural processes are seldom simultaneous but occur according to precise temporal and coordinative choreographies within and between brains. Methods with good temporal resolution such as EEG can help to identify so-called "neuromarkers" of social function (Tognoli, et al., 2007) and aid in disentangling the dynamical architecture of social brains. We have studied neuromarkers and their dynamics during synchronic interactions in which pairs of subjects coordinate behavior spontaneously and intentionally (social coordination) and during diachronic transactions that required subjects to perceive or behave in turn (action observation and delayed imitation). We examined commonalities and differences in the neuromarkers that are recruited for both kinds of tasks. We found that the neuromarker landscape was task-specific: synchronic paradigms of social coordination revealed medial mu, alpha and the phi complex as contributing neuromarkers. Diachronic tasks recruited alpha as well, in addition to lateral mu rhythms and the newly discovered nu and kappa rhythms whose functional significance is still unclear. Social coordination, observation, and delayed imitation share commonality of context: in our experiments, subjects exchanged information through visual perception and moved in similar ways. Nonetheless, there was little overlap between the neuromarkers recruited for synchronic and diachronic tasks, a result that hints strongly of task-specific neural mechanisms for social behaviors. The only neuromarker that transcended both synchronic and diachronic social behaviors was the ubiquitous alpha rhythm, which appears to be a key signature of visually-mediated social behaviors. The present paper is both an entry point and a challenge: much work remains to determine the nature and scope of recruitment of other neuromarkers, and to create theoretical models of their within- and between-brain dynamics during social interaction.

## 1. Introduction

Social neuroscience has garnered tremendous interest over the past decade, as readily appreciated from the large number of dedicated reviews (e.g. Ochsner & Lieberman, 2001; Frith & Frith, 2001; Cacioppo, 2002; Insel & Fernald, 2004; Gallese et al., 2004; Blakemore et al., 2004; Saxe, 2006; Senior & Rippon, 2006; Cacioppo et al., 2007; Behrens et al., 2009; Adolphs, 2009; Hari & Kujala, 2009; Schilbach, 2010; Farah, 2011). The entire non-invasive armamentarium of the neurosciences has been harnessed toward the goal of discovering neural mechanisms of human social behavior, for instance using EEG (Babiloni et al., 2002; Sebanz et al., 2006; Tognoli et al., 2007[a]; Lindenberger et al., 2009; Dumas, et al., 2010; De Vico Fallani, 2010), MEG (Hari et al., 1998), PET (Decety et al., 2002), fMRI (Iacoboni et al., 1999; Montague et al., 2002; Beauchamp et al., 2003; Olsson & Phelps, 2007; Izuma et al., 2008; Schilbach et al., 2010; Saito et al, 2010; Guionnet et al., 2012) and optical imaging (Suda et al, 2010; Funane et al., 2011). The stakes are high: knowledge of the brain mechanisms involved in social behaviors has tended to lag far behind knowledge of the individual brain. On the clinical side, social behaviors show intricate symptomatic and etiologic ties with a vast number of brain conditions as well as with their treatments (see table 1). As a consequence, on both basic and clinical grounds, discoveries and understanding are much needed. The neuromarker approach is motivated by the long term goal of alleviating the burden of brain disorders and is designed with this goal in mind. Table 1 indicates the relevance of social neuromarker investigation for a vast number of clinical conditions, and reveal why such studies have potentially broad implications for science and society.

Neuromarkers are important tools to describe the transient activity of the brain's functional networks during social behavior. They may appear as oscillatory patterns in electrophysiological measurements due to electrical activity that reverberates in specific brain circuits (Kelso, 1995; Buzsáki, 2006; Kelso & Tognoli, 2007; Tognoli & Kelso, 2009). Or they may appear as spatial activity patterns in imaging approaches such as fMRI. In the following, we review methodological advances developed in our laboratory and findings that followed from them within the context of experimental paradigms from social coordination dynamics. Our dynamical approach (and its corresponding dealing with continuous EEG) is geared toward the analysis and understanding of network-specific oscillatory patterns that are engaged and disengaged during social behavior. The work aims to elucidate the mapping between dynamic brain patterns and social function. To our knowledge it is a first attempt to interpret the social dynamics of brains and behaviors that is not based on average evoked potentials or average spectra and related measures, but geared toward the level of continuous brain dynamics. Similar dynamical efforts are growing quickly in the field of brain-machine interfaces, but at the present time, they have not been deployed to interpret the dynamics of social behavior. Provided the complexity of social function, it is likely that multiple routes are available for the realization of a task. The dynamical neuromarker approach also targets the characterization of such "degeneracy", so that social behavior would be better explained.

| Reference | Domain of investigation |
|---:|---|
| Compton et al., 2005 | Alcohol/drug use disorders |
| Bennett et al., 2006 | Alzheimer's disease |
| Crooks et al., 2008; Mendez et al., 2005 | Dementia |
| Segrin, 2000; Clark et al., 2003; Inoue et al., 2006 | Depression |
| Bassuk et al., 1999; Glei et al., 2005; Beland et al., 2005 | Ageing |
| Monetta et al., 2009; Poletti et al., 2011 | Parkinson's disease |
| Mundy et al., 1986; Dawson et al., 2004; Baron-Cohen & Belmonte, 2005; Hadjikhani et al., 2006; Volkmar, 2011 | Autism |
| Giovagnoli et al., 2011 | Epilepsy |
| Kirsch, 2006 | Epilepsy surgery |
| Genizi et al., 2011 | Benign childhood epilepsy |
| Yeates et al., 2007 | Children brain disorder |
| Greenham et al., 2010 | Children brain insults |
| Lezak & O'Brien, 1988; Gomez-Hernandez et al., 1997 | TBI |
| Anderson et al., 1999 | Prefrontal lesions |
| Willis et al., 2010 | Orbitofrontal cortex lesions |
| Farrant et al., 2005 | Frontal lobe epilepsy |
| Green & Phillips, 2004 | Paranoia |
| Russell et al., 2000; Couture et al., 2006; Brunet-Gouet & Decety, 2006; Schimansky et al., 2010; Varlet et al., 2012 | Schizophrenia |
| Jones et al., 2000; Meyer-Lindenberger et al., 2005 | Williams syndrome |

*Table 1: The challenge of finding social neuromarkers: evidence of altered social behavior in clinical conditions or clinical treatments.*

One of the original quests in the field of social neuroscience was directed toward discovery of "the" neuromarker of social behavior, that is, brain activity emanating from a functional network that transcends social interaction contexts-- perhaps in the form of a system of mirror neurons (Gallese et al., 2004; Uddin et al., 2007). However, from many studies including our own, it has come to pass that more neuromarkers are recruited and modulated over the course of social behavior than initially presumed. Using EEG to investigate social interactions, our findings reveal that social neuromarkers have oftentimes taken the form of oscillations in the 10Hz frequency band, a dominant frequency in the cerebral cortex and cortico-thalamic loops. In addition to portraying neuromarkers from this very active region of the EEG spectrum, we will briefly discuss the meaning and relevance of the 10Hz time scale for social behavior.

Neuromarker multiplicity has led to a number of basic questions about the functional and dynamical architecture of social brains: which major functional system do such neuromarkers attach to; how do neuromarkers differ from one another; and how do they arise and interact over the course of social transactions? Today, it seems, we are at a crossroads --having explored a sufficient task repertoire (the *behavior* side of the story) and identified a number of neuromarkers (the *brain* side)-- that it becomes possible to enquire about the integration of results and their modeling. Yet it is still early enough to say that such models will only be sketches, more like tracing a tentative path with still many elements missing from the picture and others not yet in their definitive place. The present paper is contributed in this spirit. Through methodological advances, systematic experimentation and neurobehavioral theorizing, we attempt to chart an introductory model of social brains. An enduring challenge in cognitive, affective and social neuroscience is to develop a theory of tasks (Saltzman & Kelso, 1987). This development is especially critical when dealing with dynamical models of the brain, as it may help to infer covert mental processes and determine the timing of their recruitment and dissolution. We end the present review with some ideas on how to cross this frontier in social neuroscience.

## 2. Synchronic social behaviors

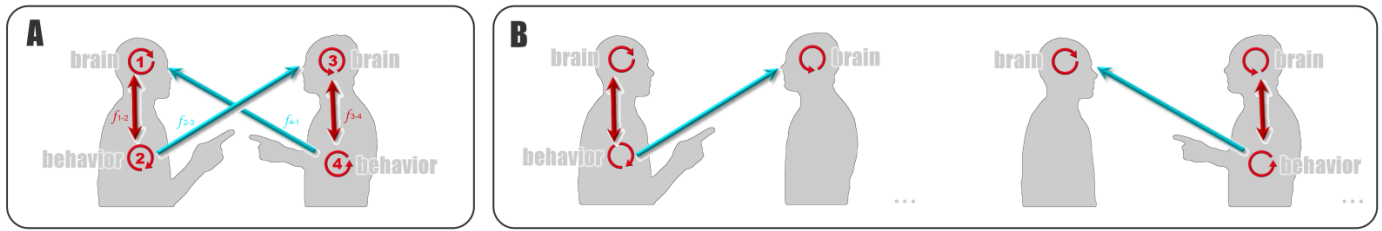

*Figure 1: diagram illustrating flow of information during synchronic (A), and diachronic (B) social interactions in a dyadic setting. Circular red arrows describe intrinsic dynamics in neural and behavioral subsystems respectively. Straight red arrows describe motor and perceptual information flows that are circumscribed to an individual; blue arrows represent information flows that cross to the other individual (social coupling). During synchronic social behaviors (A), information flows bidirectionally between all parts of the system. In contrast, during diachronic social behavior (B), only one person acts at a given time and one behavioral subsystem is disengaged. The two vignettes in B illustrate turns of behavior between the two individuals. See details in text.*

Synchronic social behaviors engage simultaneous action and perception processes. Tango dancing, choir singing, driving in traffic, executing shared-tasks such as lifting heavy furniture in tandem or performing surgery are examples of synchronic behaviors, with varying degrees of symmetry between the actions performed and the varying effector and sensory pathways that carry information across the border of individual human beings through action~perception couplings. In such interactions, information flows continuously and reciprocally between people through perceptual channels (Figure 1A, blue arrows), creating linkages at both brain and behavioral levels.

A unique characteristic of synchronic behavior is that the actions of one individual (e.g. figure 1A, 2) are readily able to modulate a partner's behavior (4) with information flowing in a reciprocal, bidirectional fashion. Information about self-produced movement is returned back to oneself and is updated based on the actions of one's partner (Figure1A, $f_{4\text{-}1}$). With both partners simultaneously engaged in such informational exchange --each continuously perturbing the other-- a system is formed that enters a kind of steady-state self-organization that exhibits rich dynamical behavior (Kelso, 1995; Sebanz, Bekkering & Knoblich, 2006; Tognoli et al., 2007[a]; Tognoli, 2008; Oullier et al., 2008; Oullier & Kelso, 2009; Riley et al., 2011). To probe this dynamics, the social coordination paradigm assesses the behavioral and neural organization of pairs of subjects as they continuously perform simple rhythmic index finger movements (extensions/flexions). Differences between behaviors produced in individual and social contexts are probed by manipulating the perceptual flow between people, switching vision of each others' action on and off with the help of an optical barrier. The great advantage of this very basic, canonical situation of behavioral coordination is that it provides explicit and continuous measures of social coupling through the dynamics of a collective variable, the relative phase (Tognoli, et al., 2007[a]; Tognoli, 2008; Oullier, et al., 2008; Oullier & Kelso, 2009; Tognoli, de Guzman & Kelso, 2011), akin to studies of bimanual (Kelso, 1984; Swinnen & Wenderoth, 2004), sensorimotor (Kelso et al., 1990; Wimmers et al., 1992) and postural coordination (Varlet et al., 2011). To study the potency of perceptual coupling and corresponding neural correlates during this spontaneous form of social coordination, we further distinguish trials during which subjects entered a state of phase-locked collective behavior from those that did not (Tognoli et al., 2007[a]).

## 3. Diachronic social behaviors

In contrast to synchronic behavior, only one participant acts at a given time in diachronic social transactions. Examples of such behavioral transactions include conversation with well-defined turn-taking, playing golf, watching a ballet dancer, imitating a person's facial expression or accent, and learning a surgical gesture by observing a demonstrator in medical school. Coupling in the system is ensured by perceptual flows to the observer's brain (Figure 1B, blue arrows), but there information flow reaches an end-point --at least momentarily until role settings are eventually modified--. As a result, information flows do not circulate continuously in the system. If all relevant influences stopped in the immediacy of perceptual exchange, this type of social transaction would seem less useful than its ubiquity suggests. However, it appears that such exchanges rely upon delayed influences –possibly buffered in the observer's brain through memory processes– and mutual social influences are therefore allowed to resume at slower time scales (see Tognoli & Kelso, *in press* for a theoretical discussion on time scales and causality in complex systems). Experimental tasks that probe such diachronic behaviors include action observation and delayed imitation. In our implementation (Suutari et al., 2010), we instructed pairs of participants to first observe then imitate index finger movements in turn, during two periods of continuous behaviors (8 sec long) separated by retention, pause and control intervals for individual behaviors. We studied social neuromarkers and their dynamics when subjects observed their partner's action, performed an action alone or under the observation of their partner, imitated the action they observed earlier, and during rest.

## 4. The neuromarker framework: finding local oscillations

From dual EEG recordings, we examined the repertoire of brainwaves (brain rhythms, periodic and aperiodic oscillations) recruited for social tasks. Brainwaves carry a 3-sided signature of underlying neurophysiological processes: (1) spatial organization (how energy is distributed over the scalp -an indirect manifestation of the originating neural structures); (2) spectral properties (the frequencies at which brainwaves operate – a manifestation of their temporal extent and affordance for interaction with other neuromarkers); and (3) functional dependency (i.e., which behavioral/mental/affective processes modulate them). In other words, analysis of brainwaves addresses the structure, dynamics and function of the brain (e.g. Bressler & Tognoli, 2006; Freeman, 2000; Başar, 2004; Buzsáki, 2006; Kelso, 1995; Kelso & Tognoli, 2007).

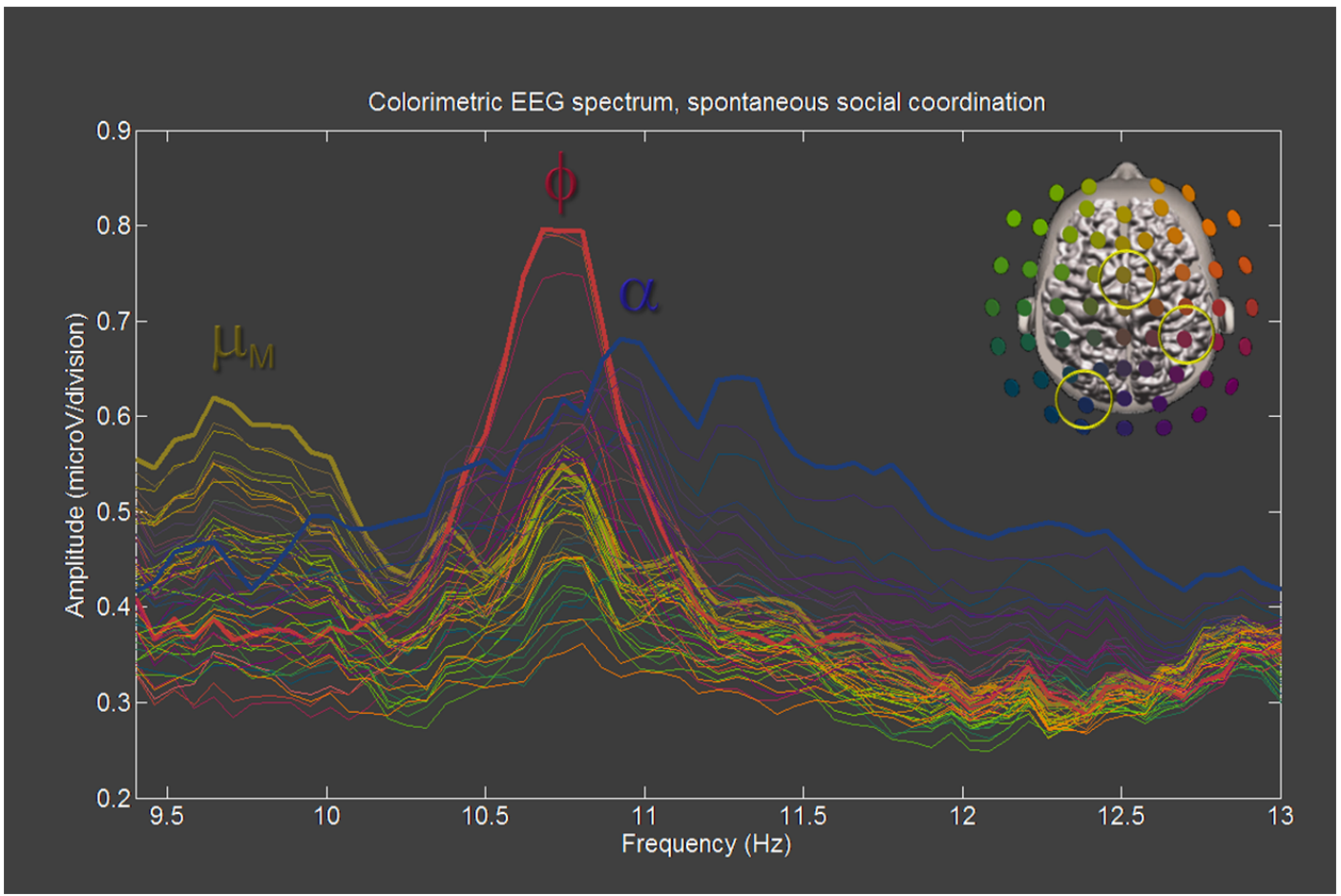

*Figure 2: Neuromarkers can be parsed using multi-electrode spectra with high spectral resolution (here bin size is 0.06Hz) and colorimetric encoding of spatial organization (following colorimetric legend shown in upper right corner). In this figure, sampled from a subject performing a spontaneous social coordination task, 3 neuromarkers are observed that include mu medial (appearing in brown color as a result of its fronto-central topography), left alpha (blue, left occipital region) and phi 1 (burgundy, right centro-parietal region). Note spectral proximity, especially for phi and alpha.*

Importantly, from the theory of coupled oscillators, it ensues that neural oscillations meant to work together need to operate on similar time scales. If the binding mechanism at play is phase- and frequency-locking or a more subtle metastability, this constraint translates into neural ensembles' operating with similar (or near integer-related) frequencies (see, e.g. Bressler & Tognoli, 2006; Tognoli & Kelso, 2009; Palva & Palva, 2007; 2012; de Guzman & Kelso, 1991: Tass, et al., 1999). As a result, spectral overlap is often present, a feature that is poorly accounted for in traditional EEG studies to the great loss of understanding brain-mind and brain-behavior relationships. For example, when examining the 10Hz band at the usual spectral resolution of ~1Hz, overlap translates into a blurred spectral and spatial differentiation of neural oscillations. More specifically, one sees an irregular-shaped peak in the spectrum, with its power distribution changing from place to place over the surface of the scalp. This amorphous view conceals a number of discrete peaks each with their own frequency and topography (such as the three peaks shown in figure 2), but so close that they may merge spatially and spectrally at low resolution. Our framework of brain coordination dynamics rests on high-resolution spectral analysis of EEG with colorimetric encoding of topography- a set of techniques that performs well at distinguishing oscillations with spectral and spatial proximity (Tognoli &

Kelso, 2009). When sufficient spatial and spectral resolution are achieved (increasing sensor density to augment spatial resolution and either increasing the amount of continuous time in Fourier analysis or lengthening the time interval artificially using zero padding to augment spectral resolution), crisp regional distributions of power do appear. Using such techniques, it is possible to measure the functional specificity of brain rhythms without the corrupting effect of other oscillations located nearby.

Although the general plan of all human brains may be the same, on a fine grain level every brain circuit is different. Hence, a neuromarker may shift slightly in frequency and topography from one subject to another. Critically, identification of neuromarkers needs to be conducted on a subject by subject basis. At this stage, interindividual comparisons are performed on parsed neuromarkers (their conditional power or/and their time course), not on the less refined picture of power distribution that is obtained from grand-average spectra (mean of all individual spectra, which again causes blurring due to spatial and spectral variations between subjects).

## 5. The neuromarker framework: brain coordination dynamics

Oscillations may be studied in average spectra (as in figure 2) and continuous time. We hypothesize that such oscillations reveal the transient activation of unique functional networks in the brain. Under such an hypothesis, it is possible to establish a time-course describing the engagement and disengagement of brain networks. The latter coexist with another timeline of descriptors, namely one that refers to the brain's functional organization at the level of behavior, perception, cognition and volition (section 6 below). The challenge for social neuroscience (and for neuroscience in general) is to recognize that both neural and behavioral/cognitive levels may be characterized in terms of their dynamics and that dynamics offers a means by which to relate them (Buzsáki, 2006; Kelso, 1995; 2012).

Neuromarker dynamics can be probabilistically approached using wavelet analysis (see e.g. Tognoli et al., 2007[a]; Suutari et al., 2010) within the spatio-spectral domain identified from a 'static' neuromarker approach (Figure 2). This provides a picture of the brain in which macroscopic ensembles fluctuate smoothly in amplitude over time, an imperfect but heuristic means to explore macroscale neural dynamics. The wavelet approach is heuristic in the sense that following selection of the right electrode and frequency band for a neuromarker of interest, it tends to maximize the correspondence between signal power fluctuations and the genuine time course of a functional process. Fundamentally, the inverse problem prevents one from identifying source dynamics solely on the basis of information from scalp recordings. As a result, electrode-based wavelet approaches (and related methods) are far from perfect. Since a number of distant neural ensembles contribute to the scalp signal in the same neighborhood, there is no guarantee that a unique neural ensemble is tracked continuously by monitoring power at selected electrodes. Rather, electrode power is determined by a number of neural ensembles in turn. A much more precise approach includes segmentation and classification of transient spatiotemporal patterns and analysis of their coordination dynamics (Tognoli & Kelso, 2009; Benites et al., 2010; Tognoli et al., 2011; Fuchs et al., 2010; and figure 3, see also Lehmann et al., 2006), to be followed by reconstruction of their source dynamics (Pascual-Marqui et al, 2002; Murzin et al., 2011). Such methods provide a picture in which sources are intermittently on and off. As discussed in (Tognoli & Kelso, 2009), we are less interested in power/amplitude quantifications (which are inappropriate measures of neural source strength in the first place), than with the lifespan of large scale patterns (duration and recurrence) and their dynamical interaction with other neural ensembles (e.g. phase relationships within patterns; vicinity of other patterns that entertain causal precedence and consequence). In our approach, all such dynamical attributes are scrutinized in terms of their possible functional significance.

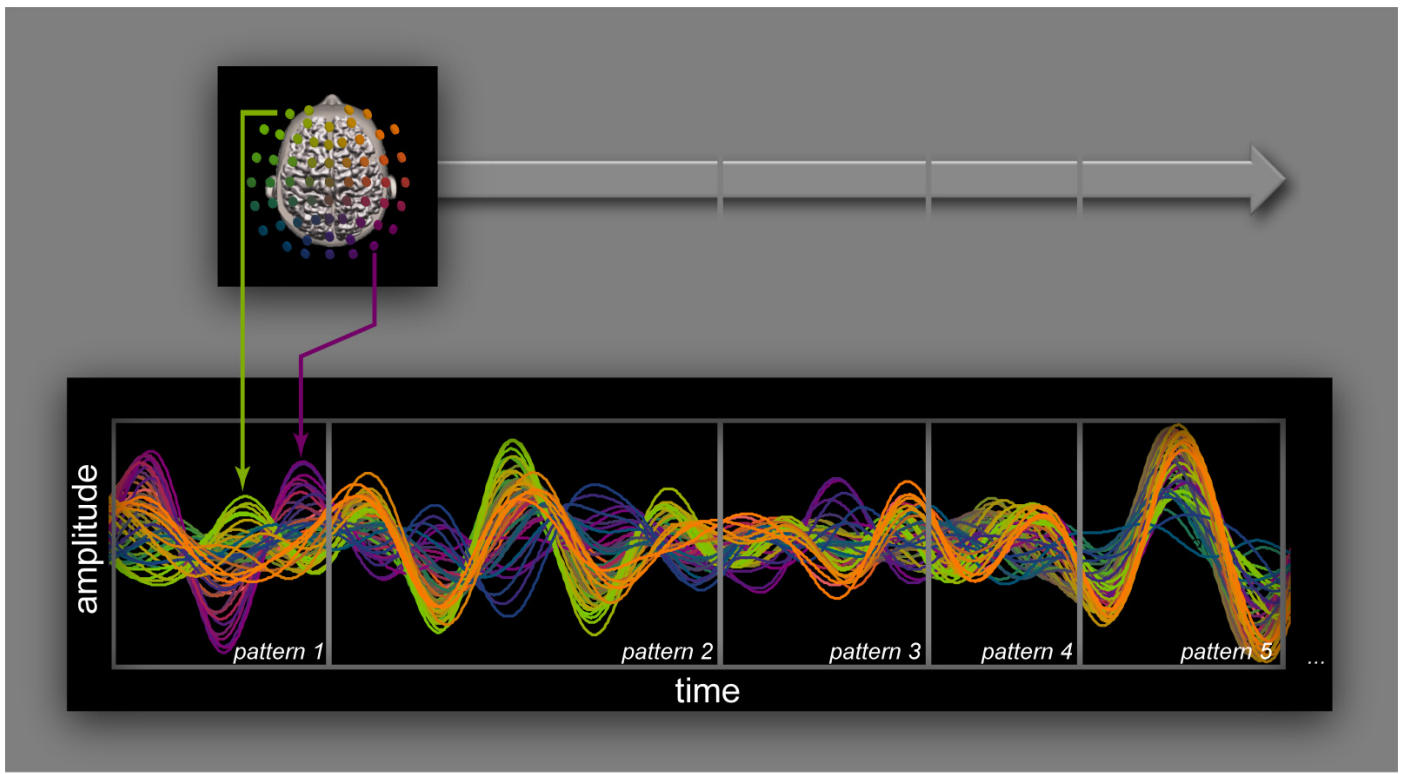

*Figure 3: A sequence of 5 patterns segmented from continuous, band-pass filtered EEG. Filters (7-13.5Hz) were set to retain activity in the 10Hz range, a prominent feature of human waking EEG. Patterns were segmented manually by two trained examiners who analyzed the spatiotemporal evolution of phase aggregates. Results were later confirmed using an automatic segmentation algorithm. Each pattern inside the grey frames is best explained by the transient organization of a few macroscopic ensembles that interact through phase-locking or metastability. For instance, the first pattern shows phase aggregates that are sugggestive of one gyral and one sulcal source (green and magenta arrows respectively; source estimation would provide some indication on their cortical origin). Short-lived configurations tend to succeed one another (e.g. magenta phase aggregate ends with the edge of the first box, giving way to new phase aggregates in the second grey box). Putatively, this organization provides support for ongoing functional processes. Note that such neural organization in the 10Hz frequency band sustains transient patterns with a typical duration of one to two hundred milliseconds, a crucial time-scale for human behavior, both individual and social.*

## 6. The neuromarker framework: functional inferences

Inferences about brain~behavior correspondences (a temporal puzzle, see figure 4) represent a key challenge that must be overcome in order to achieve adequate explanatory models of social brains. The rich phenomenological language of human behavior and cognition has been developed over centuries of scholarly enquiry, accelerated in recent decades due to the thrust of cognitive (neuro)science. We postulate that the functional language of human behavior (e.g. sociocognitive and affective processes) maps onto discrete neural patterns - those that can be captured from segmentation of continuous EEG (see section 5 above). Due to the convergent~divergent connectivity of the brain, the mapping is likely to be degenerate: the same output pattern may be produced by a number of different interacting brain structures, and alternative pathways between neural structures are capable of producing functionally equivalent cortical patterns (Edelman & Gally, 2001; Kelso, 2012; Tononi, 2010)--the key signature of self-organized synergies (Kelso, 1995). The empirical challenge then becomes one of matching temporally inferred functional processes and observed brain patterns (figure 4, left). Such inference is guided by the study of neuromarkers (as in figure 2), and neuromarker dynamics (as in figure 3). A sound strategy consists of meta-analyses: after a neuromarker has been revealed through the study of multiple tasks and experimental manipulations, it should become possible to narrow down its functional significance more precisely, thereby separating its true functional meaning from sporadically co-varying effects.

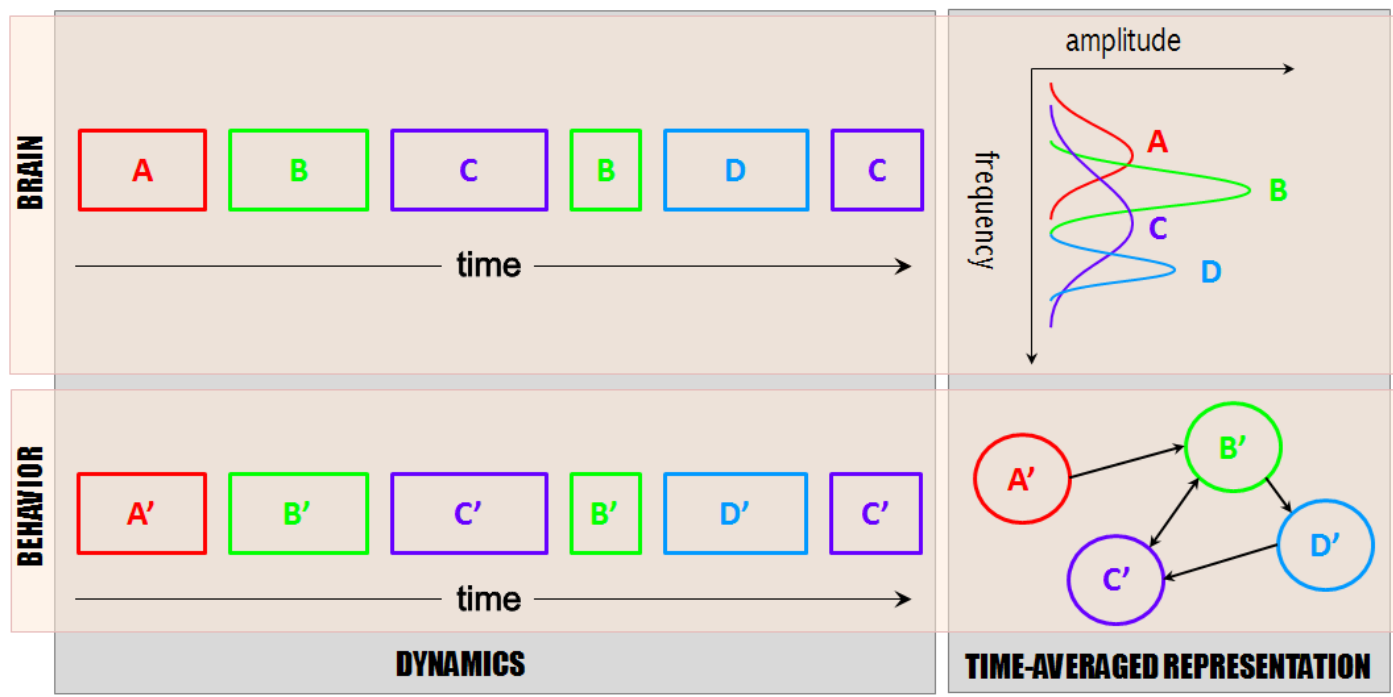

*Figure 4: Dynamical descriptions of brain functional networks (top left) and inferred functional processes (bottom left), along with their time-averaged representation (power spectrum on the upper right; functional graph on lower right). For simplicity, only one frequency band is represented (say, 10Hz), and only one process at a time (i.e. no network interaction). In reality, multiple frequency bands (and associated functional processes) occur at the same time. Typically, networks are co-activated and exhibit transient interactions. The goal of functional inference is to identify the functional processes (bottom rectangles) that match spatiotemporal patterns of brain activity (top rectangles) and their temporal footprints, so that correspondences between brain and behavior can be uncovered. Though simplistic, a translational language along these lines would propel our understanding of social brain functions and lead the way toward explanatory models.*

Difficulties lie in the fact that (1) theories of tasks are seldom based on explicit, observable quantities; and (2) such descriptions, despite their ready reduction into serial models, are not grounded in a dynamical framework that allows one to establish unambiguous time addresses for the engagement and disengagement of functional processes. A place to begin such an endeavor is with functional processes that have explicit temporal footprints, as in our social coordination paradigms. Time-averaged neuromarkers (obtained from the methodology spelled out in section 2) and their reactivity also provide tractable material that may lead to establishing neuro-functional relationships (see table 2).

Descriptions of behavior and cognition are especially fruitful for slower and more global functional processes, the time-scale of which was amenable to observational and experimental tools of earlier times. In contrast, faster processes have not systematically received distinct names and descriptions. Short-lived patterns that are uniquely tracked with dynamic brain imaging techniques such as EEG and MEG may hold keys to advancing understanding of social behavior. Understanding causal chains of neuro-functional processes at faster time-scales may be one of the most valued advances that social neuroscience can make.

## 7. Neuromarker commonalities and differences

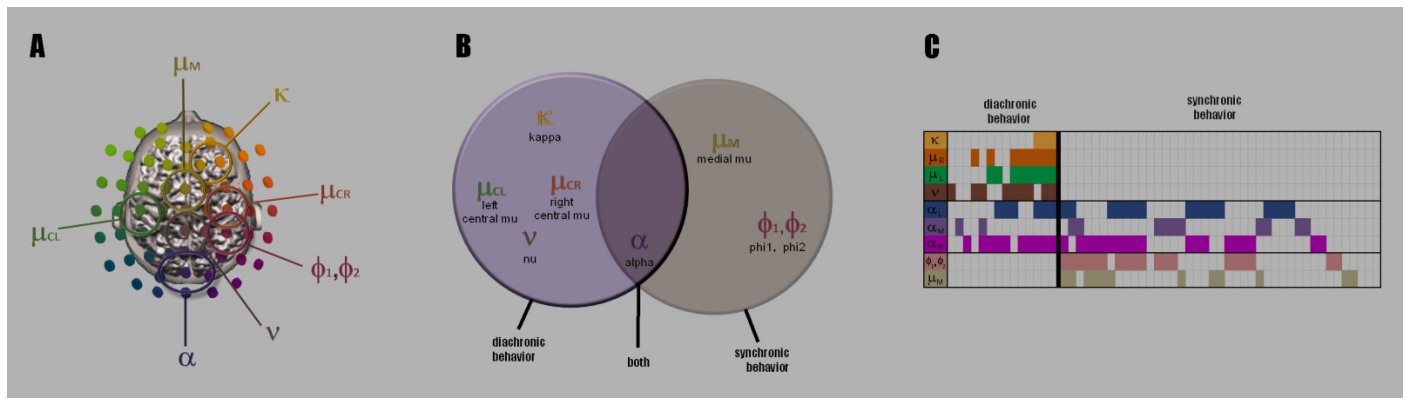

*Figure 5: overview of neuromarkers contributing to social behavior, with their scalp topography (A), a Venn diagram of their recruitment in our studies of synchronic and diachronic social behavior (B), and a meta-analytic table of their inter-individual variability (C). Neuromarker location in (A) indicates sensor carrying highest power on the scalp, keeping in mind that this does not imply regional homology with underlying cortical structures. Each column of (C) specifies a subject enrolled in our experiments of social behavior, each row corresponds to a neuromarker. When a neuromarker was detected in a subject, its cell is marked with a color, else, it is left blank. Note empty sectors in the lower left and upper right sectors that suggest specific neuromarker landscape for the two types of social behaviors.*

The repertoire of neuromarkers modulated during our social tasks is summarized in figure 5. During synchronic social behavior, a set of neuromarkers was recruited that included the alpha rhythm, the phi complex and, when interaction was spontaneous, a medial mu rhythm (Tognoli et al., 2007[a]; Tognoli et al., 2007[b]). During diachronic social behavior, alpha was also observed, but mu medial and the phi complex were not detectable. In addition, left and right central mu appeared as did two newly described nu and kappa rhythms (Suutari et al., 2010). The spatial, spectral and functional properties observed for these rhythms in our samples of subjects are reported in table 2. Keeping in mind the high-resolution spectral analysis that we have implemented, accuracy of estimation is aligned with the spectral resolution of our coarsest dataset, that is, 0.1Hz. The data presented in Table 2 are group results obtained from the samples of subjects that have participated in our studies (peak frequency describes the arithmetic mean of the samples; electrode location refers to the mode). Of course, large populations would be helpful to establish robust normative properties of neuromarkers (something that at this time, we forgo in favor of smaller, discovery-based studies). In table 2, we outline the spatial, spectral and functional properties as a starting point to identifying new neuromarkers and with the aim of helping other scientists who share similar goals.

|  | Neuromarker | Peak Frequency | Topography | Task dependence |
|---|---|---|---|---|
| Only in synchronic tasks | Mu medial | 9.1Hz (1.1) | FCz | - recruited by tasks of spontaneous coordination<br>- suppressed during social interaction<br>- rebounds during intrinsic, self-produced movement |
|  | Phi complex | 10.9Hz (0.9) | CP4 | - phi 1 increases in coordination tasks when subjects avert synchronization with each other<br>- phi 2 increases when subjects synchronize with each other<br>- phi complex recruitment and modulation are strongest during intentional social coordination |
| Both task types | Alpha | 9.9Hz (0.9) | PO7 (left), PO8 (right), POz (midline aggregate) | - increased by drowsiness<br>- decreased by visual input<br>- decreased further by vision of the partner's movement (larger decrease than with non-social stimuli)<br>- further decreased if partner's movement is more variable |
| Only in diachronic tasks | Left and right central mu | Left: 10.6Hz (0.9)<br>Right: 11.4Hz (0.8) | C3 (left), and C4 (right) | - left and right mu depressed during self movement; rebound at self-movement arrest<br>- do not decrease systematically during social observation<br>- exhibit dynamic aftereffects (suppression, rebound) at cessation of observed movement.<br>- right mu vanishes tonically when people memorize observed behavior (right hand movement) |
|  | Nu | 10.1Hz (1.1) | CPz | - decreases during self-movement<br>- increases when self-movement is performed in view of another person |
|  | Kappa | 11.2Hz (0.7) | FC2 | - tendency to decrease any time either partner performs a movement in view of the other |

*Table 2: summary of spatial, spectral and functional properties of neuromarkers involved in synchronic and diachronic social behavior. The data presented are group results obtained from the samples of subjects that have participated in our studies. Peak frequency (measured from high-resolution spectra) describes the arithmetic mean of the samples with standard deviation in parenthesis. The electrode reported in column topography refers to the mode (electrode most frequently observed across subjects to bear largest spectral energy, electrode names are according to 10 percent system, Chatrian et al., 1985); recording was performed with linked-mastoid reference. 'Task dependence' refers to conditions in which power is modulated, a precursor to inferences about function.*

The only neuromarker that transcended both synchronic and diachronic social behavior was the alpha rhythm, a neuromarker associated with visual attention (Mulholland, 1972; Klimesch et al., 1998; Palva & Palva, 2011). All of our studies revealed that vision of the partner substantially reduced alpha power. With its separation of social- and self-behaviors in distinct experimental phases, our study of action observation further allowed us to show that alpha fluctuated with the complexity of behavioral information acquired about the partner. In Suutari et al. (2010), single trial alpha power was low when observers were exposed to finger movements with high cycle to cycle variance. By contrast, alpha increased with more regular movements. Put another way, the individual brain's alpha rhythm appears to be a pertinent measuring instrument of the complexity embedded in interpersonal information flows (see also Müller et al., 2003 for related account in non-social visual perception).

A social interaction exists only if social partners acquire information about each other (see blue arrows in figure 1). Our results suggest that the alpha rhythm is a key neuromarker of visually-mediated social behavior (putatively, social transactions mediated by other sensory channels would have their own signatures, see, e.g., Pineda, 2005 for candidates). Alpha modulation is often overlooked in EEG/MEG studies of social interaction in favor of mu rhythms. We suggest however that alpha's sensitivity to informational exchange between partners, its large amplitude in human EEG and robust presence in most subjects makes it an important neuromarker of social behavior (see section 4 for strategies to disambiguate alpha, mu and other spectrally similar neuromarkers). Furthermore, in visual detection tasks, it has been shown that alpha suppression is spatially informative, with attention to the right hemifield depressing specifically left alpha rhythm and vice-versa (Worden et al., 2000; Sauseng et al., 2005). Such lateralization could be useful to disentangle self and social attention in experimental designs that carefully manipulate the spatial arrangement of self and other-- with the potential outcome that roles in social interactions could be quantified as a function of the spatial deployment of attentional resources. Moreover, interindividual variation in alpha suppression could reveal the extent of social engagement and task-related social affinities, with consequent applications to a variety of domains relevant to human social behavior.

## 8. Toward dynamical models of social brains

As we observe many neuromarkers and their intermittent dynamics in dual-EEG recordings (section 5), we are led to question their spatiotemporal organization –how the functional processes that participate in social behavior are orchestrated. Until now, at the largest scale of complete dual-EEG experiments, we have achieved either a static neuromarker description (as in section 4), or a probabilistic description of their dynamics using wavelet analysis on selected frequency bands and spatial sites (e.g. Tognoli et al., 2007[a]; Suutari et al., 2010, see discussion in section 5). Based on theoretical and methodological work (Tognoli & Kelso, 2009), we have also started to study the dynamic patterns of dual-EEG (see figure 3 and text thereafter) on particularly interesting aspects of social behavior such as the loss or establishment of coordinated action. The first stage of this analysis is a segmentation of continuous (band-selected) EEG. We have implemented either a manual analysis of the oscillations' phase, frequency and topography (Benites et al. 2010), or an automatic segmentation method examining the eigenvalue tradeoff between two principal modes of the EEG power envelope derived from a rotating wave approximation (Fuchs et al., 2010). The result of both approaches is to parse each participant's EEG into a sequence of dynamic patterns (see figure 3). This sequence is then matched to an estimation of the time course of inferred functional processes (Figure 4), with the goal of connecting their dynamics. This framework extends our earlier efforts that found a tight connection between behavioral and neural dynamics once an appropriate space of collective variables has been identified. Spatiotemporal measures of brain activity tracked kinematic measures of sensorimotor coordination both empirically (Kelso et al., 1998) and in a theoretical model of the underlying neural field dynamics (Fuchs, Jirsa & Kelso, 2000).

As more and more insights into the function of neuromarkers becomes available, it should become possible to solve the temporal puzzle of brain~behavior as presented in figure 4. When that point is reached, we will be able to draft dynamical models of social processes at the combined levels of brain and behavior and to study their variation in different situations (e.g., social skill development, disease, effects of pharmacological treatment, etc.).

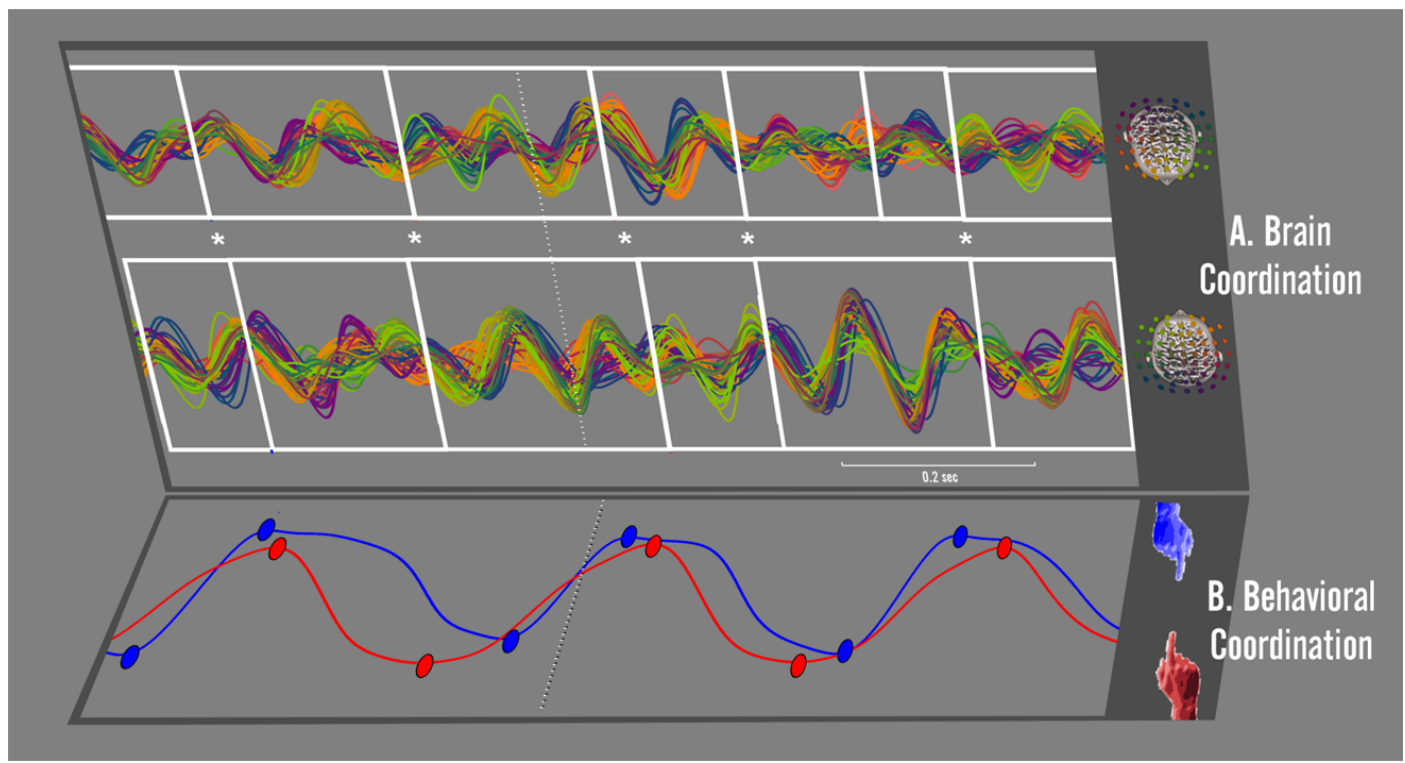

*Figure 6: Synchronized patterns between brains. Continuous dual-EEG is shown in the 10Hz frequency band for a pair of interacting subjects in A, with electrode signals encoded using the colorimetric legend shown on the right (EEG from one subject on top, the other on bottom). Changes in spatiotemporal organization of brainwaves were determined by two trained examiners who were blind to the associated behavioral variables. A manual segmentation was performed separately on each subject's EEG. Transitions are marked by successive white frames, following the method outlined in section 5 and figure 3. In this sample trial, subjects were instructed to coordinate finger movements inphase (see red and blue movement trajectories of right index fingers in B). The dashed line bissecting plots A and B indicates the moment at which subjects successfully coordinated their behavior (with the movements' relative phase exhibiting a sudden phase transition to inphase – not shown). The entire temporal window displayed is about one second long and relates to the intentional transition process from independent to coordinated behavior. In this window, the transition between subjects' brain patterns reveals strong tendencies for coincidence (see series of asterisks in A, cueing temporal proximity of each subject's brain pattern transitions). Note that the dynamic patterns of each participant's brain activity have distinct spatial, spectral, and phase organization. Neural transitions are coupled, but not the spatiotemporal neural patterns located between them.*

In the preceding, we have examined collective behavior and its relation to brain activity, but only a single brain at a time. With social neuroscience born from cognitive neuroscience, there is a temptation to segregate the neural activity of participants to fit the existing framework of single-brain neuroscience. A true social neuroscience, however, will only realize itself when it fully integrates neural activity of every participant in a common analysis scheme. Efforts to do so have been undertaken by collecting synced records of brain activity from multiple people (e.g. dual-EEG: Tognoli et al., 2007[a]; or fMRI hyperscanning: Montague et al., 2002) and by formulating novel analysis frameworks that combine the neural dynamics from multiple subjects (Lindenberger et al., 2009; Dumas et al., 2010; Dodel et al., 2011; see also Tognoli, de Guzman & Kelso, 2011). With brains chock full of oscillations that are coupled between people through inter-personal perceptual flows, a straightforward hypothesis is that oscillations enter collective states of phase-locking and frequency coupling between the brains of interacting partners--a hypothesis that has been pursued by ourselves and others (e.g. Lindenberger et al., 2009; Dumas et al., 2010; see also Funane et al., 2011, for related hemodynamic account). Our research has yet to uncover unambiguous evidence of phase-locking between the brains of people as they engage in social behavior. Moreover, our longstanding theoretical inclination is toward metastable coordination, where tendencies for integration coexist with tendencies for segregation (e.g. Kelso, 1995; Kelso & Tognoli, 2007). The reason we suspect that we don't observe phase synchrony is that at the level of dynamic patterns (and in the frequency bands examined, especially around 10Hz), limited symmetry exists between the instantaneous networks formed in each person's brain (see example figure 6). However, in applying the aforementioned segmentation methods to social coordination tasks, we encountered evidence of another, less expected mechanism of coupling between brains (Benites et al., 2010; Fuchs et al., 2010). On one hand, each subject's neuro-functional activity was distinct (compare upper and lower white frames in figure 6A, and note patterns' lack of correspondence in topography, frequency and phase), but on the other hand, the moment at which those patterns changed in each partner coincided (note temporal coincidence of white frames' edges marked with asterisks in figure 6). In other words, it was not the oscillatory neural activity proper that was synchronized between people but rather the underlying temporal structure of their recruitment and dissolution. An analogy to such inter-brain coordination is a group of musicians, each playing different notes yet achieving a harmonious outcome by following the same tempo—without, of course, a conductor (see Kelso & Engstrom, 2006, p.93). We hypothesize that this mechanism of inter-brain coordination springs from the very weak coupling engendered by perceptual flows (i.e., weaker than connectivity-based information flows within brains). We further speculate that this weak coupling promotes the emergence of complexity in social interaction (Tognoli, de Guzman & Kelso, 2011).

## 9. Summary & conclusions

Social neuroscience is a young discipline. Accordingly in this review we have focused more on finding the right questions than providing definitive answers about the functional and dynamic architecture of social brains. Our aim is to establish a comprehensive framework to study the dynamics of brains as they evolve through successive phases of social interaction. Such a dynamical framework seems necessary if we are to understand normal and pathological social function. Using a novel set of techniques, a number of neuro-functional signatures of social behavior have been uncovered, each providing a specific topography and frequency, and each affording measurements that extend to the upmost temporal precision of continuous brain dynamics. We have drafted some tentative directions for functional inference on newly discovered and lesser known neuromarkers, keeping in mind that more information is needed to converge upon solid interpretations.

Social behavior is grounded in perception~action coupling: in the absence of action from an individual, there is no information flow to carry to another's brain. Without sensitivity to this information by a "receiver's" perceptual system, there can be no social interaction either. We stressed the primacy of information flows across individuals, and we showed their fundamental importance for visual attention, an aspect that we believe, has received insufficient scrutiny in social neuroscience.

We examined interpersonal perception-action coupling from the standpoint of the relative phase between individuals (simultaneous or diachronic action~perception). Of course, what we describe as synchronic and diachronic behaviors are limit-cases of a continuum of social circumstances that varies systematically with the phase of each participant's action. Yet, heuristically, this taxonomy proved useful in revealing little overlap between respective neuromarker landscapes. At several levels of temporal precision (e.g. across tasks, through average activity over trials, and through instantaneous activity), we emphasized the complex reorganization of endogenous brain networks (phase transitions and bifurcations) leading to the different phases and facets of social behavior.

From the multiplicity of functional processes, and from our findings that the underlying neuromarkers tend not to arise simultaneously, we have begun to enquire about their engagement and disengagement over the course of social interaction, a step that we hope will help refine functional (dynamical) modeling. In our opinion, much work remains to unravel the neural choreography of cognitive, affective and behavioral processes that participate in social behaviors and to embed them in theoretical/computational models of social brain function. Keys to future progress lie with the study of neuromarker dynamics, a step that will lead to modeling the neuro-functional architecture of social brains.

Already, the present dynamical approach to social brains has revealed some unique coordinative mechanisms that truly relate to social neuroscience (as opposed to a generalization of cognitive neuroscience to social tasks). That is, with the help of the dynamical framework presented in section 5, we have encountered preliminary evidence that spatiotemporal patterns of brain activity tend to switch in synchrony in pairs of subjects that establish or dissolve behavioral coordination (Benites et al, 2010; Fuchs et al., 2010). These synchronized transitions happened even as one subject's neural activity differed from that of the other. This finding reveals once more that the interplay of integrative and segregative tendencies within (and now between) brains is a powerful mechanism of nature to enhance system complexity (Edelman, 1999; Sporns, 2003; Kelso, 1995; Kelso & Tognoli, 2007; Tognoli & Kelso, *in press*). It is at the level of multiple brains and multiple behaviors, within a complex systems framework, that dynamical models of social function are likely to be ultimately formulated.

**Acknowledgments**

We acknowledge the contribution of HBBL group members who took part in the work that led to this review, especially Gonzalo de Guzman, Julien Lagarde, Daniela Benites, Benjamin Suutari, Seth Weisberg, William McLean and Armin Fuchs. We are grateful to the agencies that supported the theoretical, methodological and empirical work of our Social Neuroscience research program, and especially, NIMH (MH080838), NSF (BCS0826897), the US ONR (N00014-09-1-0527), and the Davimos Family Endowment for Excellence in Science. JASK was also supported by the Chaire d'excellence Pierre de Fermat.